\documentclass[aps,longbibliography,showpacs,twocolumn,superscriptaddress]{revtex4-1}
\usepackage{amsmath,amssymb,amsfonts,bm}
\usepackage{graphicx}
\usepackage{epstopdf}
\usepackage{dcolumn}
\usepackage{mathrsfs}
\usepackage{bbold}
\usepackage{dsfont}
\usepackage{float}
\usepackage[colorlinks=true,linkcolor=blue,citecolor=green, urlcolor=blue,bookmarks=false]{hyperref}

\usepackage{tgtermes}
\begin{document}

\title{Topological excitonic corner states and nodal phase in bilayer quantum spin Hall insulators}
\date{\today }
\author{Zheng-Rong Liu}
\affiliation{Department of Physics, Hubei University, Wuhan 430062, China}
\author{Lun-Hui Hu}
\affiliation{Department of Physics, the Pennsylvania State University, University Park, PA, 16802, USA }
\author{Chui-Zhen Chen}
\affiliation{Institute for Advanced Study and School of Physical Science and Technology, Soochow University, Suzhou 215006, China.}
\author{Bin Zhou}
\affiliation{Department of Physics, Hubei University, Wuhan 430062, China}
\author{Dong-Hui Xu}\thanks{donghuixu@hubu.edu.cn}
\affiliation{Department of Physics, Hubei University, Wuhan 430062, China}

\begin{abstract}
Interaction induced topological states remain one of the most fascinating  phenomena in condensed matter physics.
The exciton condensate has recently sparked renewed interest due to the discovery of new candidate materials
and its driving force to realize exotic topological states.
In this work, we explore the exciton order induced high-order topology in the bilayer quantum spin Hall insulators, and find that the topological excitonic corner states can be realized by tuning the gate and magnetic field.
When an in-plane Zeeman field is applied to the system, two or four excitonic boundary-obstructed corner states emerge in the bilayer system for distinct possible $s$-wave excitonic pairings.
Besides, we also find a two-dimensional excitonic Weyl nodal phase, which supports flat band edge states connecting the bulk Weyl nodes.
\end{abstract}

\maketitle

\emph{\color{magenta}Introduction.}---Excitonic insulator~\cite{mott1961transition,knox1963,Keldysh1964,Jerome1967}, predicted in the 1960s, is an unconventional insulating state formed by the condensation of excitons (bound electron-hole pairs) stemming from the Coulomb interaction between electrons and holes in the conduction and valence bands, respectively.
An excitonic insulator is analogous to a Bardeen-Cooper-Schrieffer superconductor that results from the condensation of electron Cooper pairs developed near the Fermi surface.
Candidate materials for excitonic insulators previously studied include quantum well bilayers~\cite{fogler2014high}, quantum Hall bilayers~\cite{li2017excitonic,liu2017quantum}, Ta$_2$NiSe$_5$~\cite{lu2017zero,werdehausen2018coherent,Wakisakaprl2009,Kanekoprb2013,Sugimotoprl2018,Mazzaprl2020}, and 1$T$-TiSe$_2$~\cite{Kogar1314,Cercellierprl2007,Kanekoprb2018,Chenprl2017}.
After the discovery of topological insulators, excitonic insulators with topologically non-trivial properties have been investigated intensively~\cite{Seradjeh2009,Hao2011,Cho2011,Tilahun2011,Efimkin2012,Budich2014,Pikulin2014,Chenprb2017,Huprl2018,zhu2019gate,Huprb2019,wang2019prediction,varsano2020monolayer,Blasonprb2020,Perfettoprl2020,SunPRL21}.
Particularly, a few representative topological exciton condensates, including the time-reversal invariant $s$-wave topological exciton condensate and the time-reversal breaking topological exciton condensate with $p$-wave pairing, have been predicted in HgTe/CdTe~\cite{Budich2014} and InAs/GaSb~\cite{Pikulin2014} quantum wells, respectively.
Importantly, the evidence for the existence of a topological excitonic insulator state has been reported experimentally in InAs/GaSb quantum wells~\cite{Dunatcomm2017,yu2018anomalously}.

Recently, the concept of topological insulators was generalized, and a novel topological phase of matter dubbed higher-order topological insulator~\cite{Zhang2013PRL,SlagerPRB2015, SlagerPRB2015,Benalcazar2017Science,Langbehn2017PRL,Song2017PRL,Benalcazar2017PRB,Schindler2018SA} was established.
Compared to the well-known topological insulators, higher-order topological insulators exhibit an unusual form of bulk-boundary correspondence.
For instance, a second-order topological insulator in two dimensions exhibits topological gapless boundary states at its zero-dimensional boundary corners, in contrast to a conventional two-dimensional~(2D) first-order topological insulator which features topologically protected gapless states at its one-dimensional edge.
So far, higher-order topology has been explored in various physical systems ~\cite{,Ezawa2018PRL,Ezawa2018PRL2,Geier2018PRB,Khalaf2018PRB,Ezawa2018PRB1,Kunst2018PRB,
	Guido2018PRB,Franca2018PRB,You2018PRB,Kooi2018PRB,
	Luka2019PRX,Feng2019PRL,Fan2019PRL,Zhijun2019PRL,Sheng2019PRL,
	chenruiprl,Varjas2019PRL,hua2020prbr,Roy2019PRB,Benalcazar2019PRB,Rodriguez2019PRB,YanPRL2018,Wang2018PRL,Zhuprb2018,Dwivedi2018PRB,Yuxuan2018PRB,Tao2018PRB,Volpezprl2019,Liu2019PRL,Yan2019PRL,Yan2019PRB,Fulga2019PRB,ZhangRX19PRL1,ZhangRX19PRL2,ZhangRX20PRL,ZengPRL19,Peterson2018Nature,Xue2018NM,Ni2018NM,Schindler2018NP,Imhof2018NP,Noh2018NPh,
	Zhang2019NP,Kempkes2019NM,Mittal2019NPh,Hassan2019NPh,Lee2020npjQM,PhysRevLett.123.196402,PhysRevLett.122.256402,Yue2019NP,Renprl,Wuprl}, and particularly Majorana zero-energy corner modes~\cite{YanPRL2018,Wang2018PRL,Zhuprb2018} were found in superconducting systems.
Yet, the topological exciton condensates known to date belong to the first-order topological phases, and the study of higher-order topology in exciton condensates is still lacking.
Given the similarity of excitonic insulators to superconductors, it is highly desirable to explore higher-order topology of exciton condensates in a realistic system.

\begin{figure}[hptb]
	\includegraphics[width=8cm]{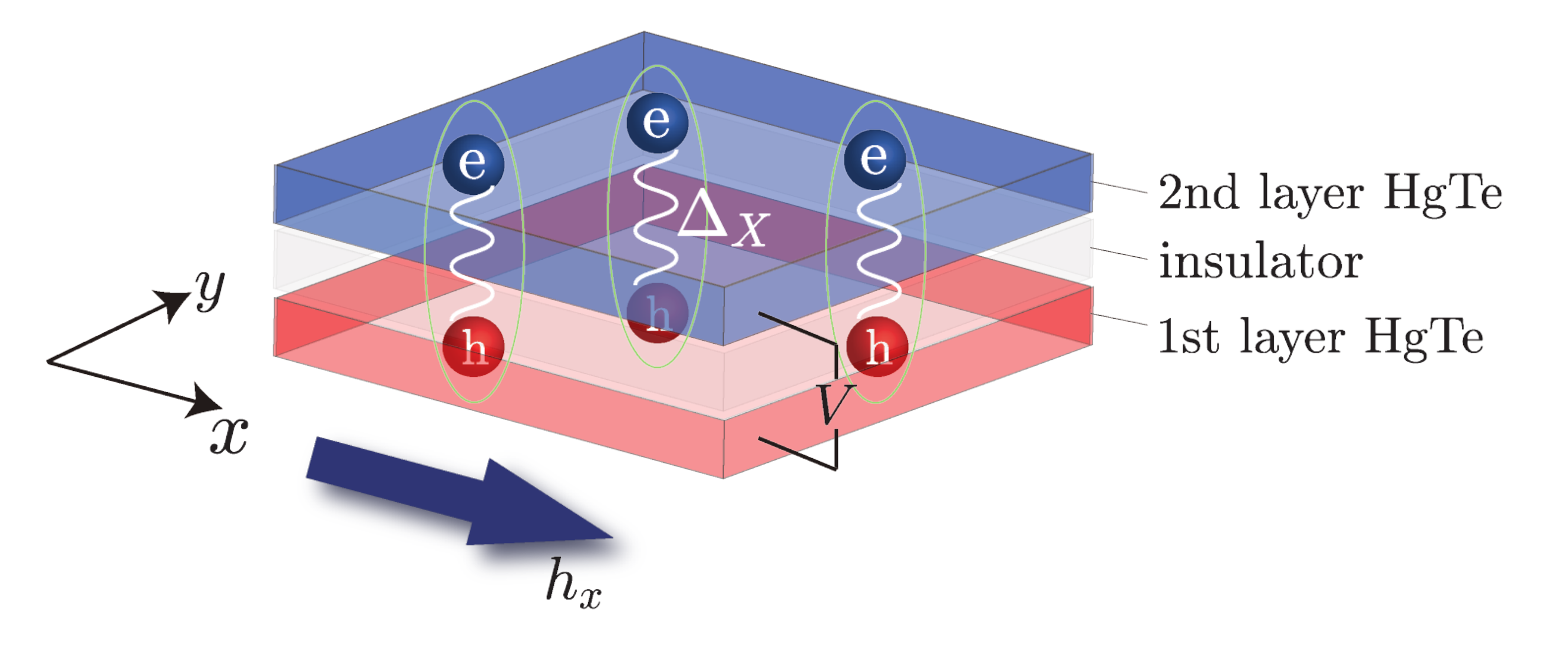} \caption{Schematic illustration of the two coupled HgTe quantum wells with gating (electric bias $V$) under the applied in-plane Zeeman field $h_x$. Electrons and holes (marked by blue and red spheres) residing on two opposing layers are spatially separated by an insulator layer. The excitonic gap $\Delta_{X}$ couples electrons with holes to form an excitonic insulator through the interlayer Coulomb interaction. }%
	\label{fig1}
\end{figure}

In this work, we demonstrate that topological corner states, which are considered as a smoking-gun signature for 2D second-order topological insulators, can be realized in the bilayer quantum spin Hall insulators with $s$-wave interlayer excitonic pairings. The bilayer quantum spin Hall insulators can be constructed by two coupled HgTe/CdTe quantum wells~\cite{GusevPRB2020} shown in Fig. \ref{fig1}, where the Dirac mass can be tuned by varying the thickness of the central HgTe layers in the quantum wells.
For the bilayer system with the negative Dirac mass, four topological excitonic corner states~(ECs) are generated by applying an in-plane Zeeman field.
By making use of the $\mathbf{k}\cdot\mathbf{p}$ edge theory, we found that ECs originate from the edge mass domain walls formed by the interplay of the excitonic order and the Zeeman field. In the case of the positive Dirac mass, however only two topological ECs emerge. In addition, we show that an intriguing 2D excitonic nodal phase, which supports flat band edge states connecting the bulk Weyl nodes, could also be realized in this system.
Our study suggests that bilayer quantum spin Hall insulators can serve as a platform to host excitonic higher-order topological insulating and nodal phases.

\begin{figure}[hptb]
	\includegraphics[width=8cm]{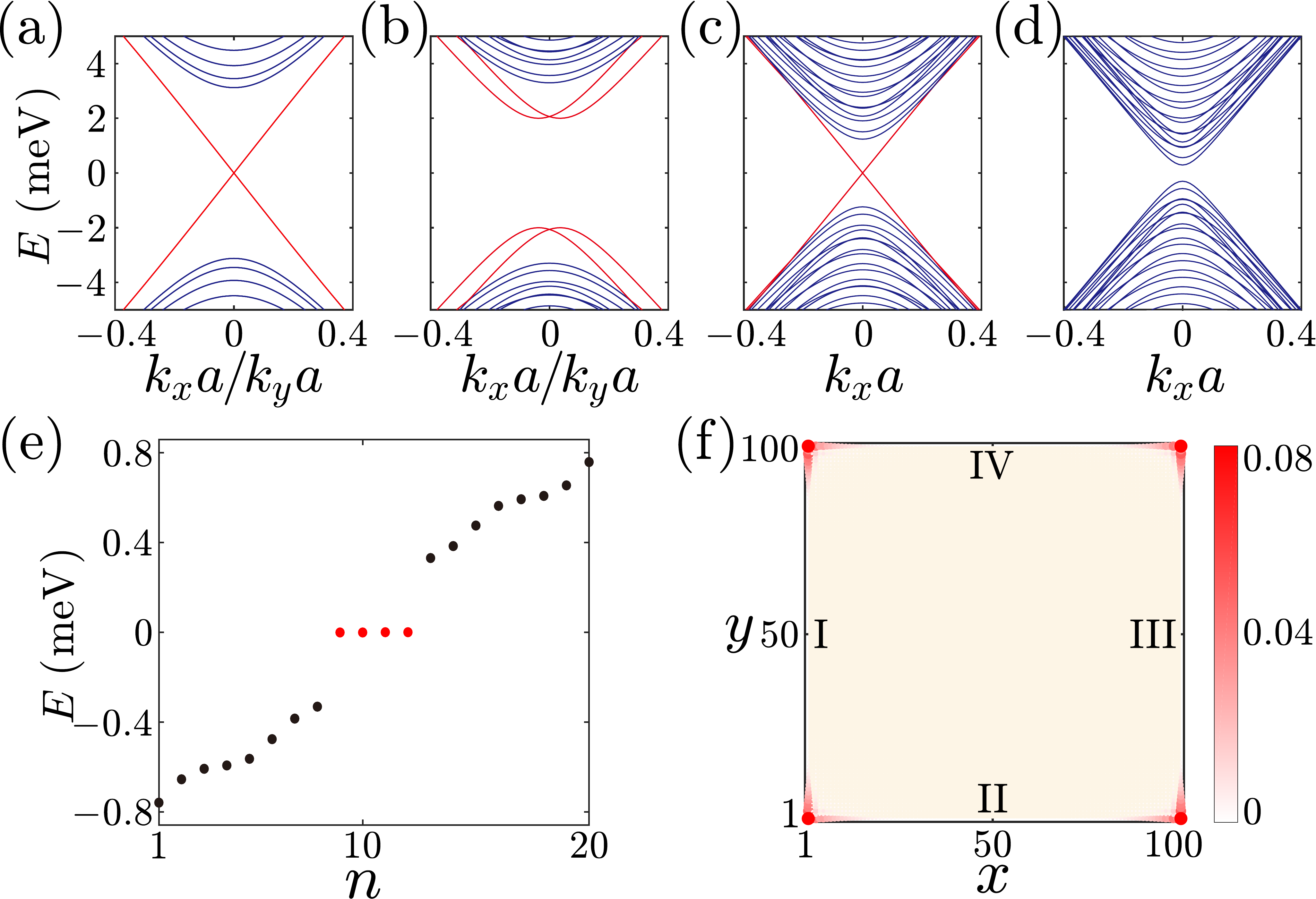} \caption{Band structure of a ribbon geometry for (a) $V=\Delta_{X}=h_x=0$, (b) $V=1$ meV, $\Delta_{X}=2$ meV, $h_x=0$, (c) $V=1$ meV, $\Delta_{X}=2$ meV, $h_x=h^c_x\approx 2.06$ meV, and (d) $V=1$ meV, $\Delta_{X}=2$ meV, $h_x=3$ meV. In (a) and (b), open boundary conditions along the $x$-direction or $y$-direction give rise to the same results. We take the open boundary condition in the $y$-direction for (c) and (d). The edge states marked by solid red lines are merged with the 2D bulk in (d). (e) Energy spectrum of an $L_x\times L_y=100\times 100$ square-shaped sample exhibiting four midgap corner states marked by the red dots. We use the same parameter values as in (d). (f) Probability distribution of corner states indicated in energy spectrum in (e). }%
	\label{fig2}
\end{figure}

\emph{\color{magenta}Excitonic corner states for the negative Dirac mass.}---
We report that a high-order topological excitonic insulator can be achieved in the bilayer quantum spin Hall insulators by tuning the bias voltage and in-plane Zeeman field.
The low-energy effective Hamiltonian of the gated bilayer quantum spin Hall insulators with an applied Zeeman field in the momentum space is given by
\begin{align}
H_{\text{QSH}}(\mathbf{k})\!=\! M(\mathbf{k})\sigma_{z} \!+\!\! A( k_x\sigma_x s_z\!+\! k_y \sigma_y) \!-\!\!\frac{V}{2}\tau_{z}+ \mathbf{h}\cdot\mathbf{s},
\end{align}
where the basis is $c_{\mathbf{k}l}^{\dag}=(c_{\mathbf{k}l\alpha\uparrow}^{\dag },c_{\mathbf{k}l\alpha\downarrow}^{\dag },c_{\mathbf{k}l\beta\uparrow}^{\dag },c_{\mathbf{k}l\beta\downarrow}^{\dag })$, $\alpha$ and $\beta$ are different orbital degrees of freedom with opposite parity, $\uparrow$ and $\downarrow$ represent electron spin, and $l=1,2$ is the layer index. $s_{x,y,z}$, $\sigma_{x,y,z}$ and $\tau_{x,y,z}$ are the Pauli matrices acting on the spin, orbital and layer degrees of freedom, respectively. $\tau_{0}$, $\sigma_{0}$ and $s_0$ are the $2\times2$ identity matrices. $M(\mathbf{k})=M-B(k_x^2+k_y^2)$, where the Dirac mass parameter $M$ determines the topological insulator phase, and $V$ is the bias potential. $\mathbf{h}$ denotes the applied in-plane Zeeman field.
When $V\!=\!0$, this Hamiltonian is exactly two copies of the Bernevig-Hughes-Zhang~(BHZ) model~\cite{bernevig2006quantum} that describes the HgTe/CdTe quantum wells. The topologically nontrivial phase of the BHZ model on a square lattice exits when $0\!<\!M/(2B/a^2)\!<\!2$ with the lattice constant $a$. We have assumed the spatial separation between these two layers to be sufficiently large so that the single-particle tunneling between layers can be neglected. Note that, the model parameters are dependent on the thickness of the quantum wells. In subsequent calculations, the following parameters remain unchanged, $A\!=\!275~\rm {meV \cdot nm}$, $B\!=\!-1300~\rm {meV \cdot nm^{2}}$, and the lattice constant is $a\!=\!20~\rm {nm}$~\cite{Budich2014}. The results remain valid when the parameters vary. For our purpose, we set the Dirac mass $M\!=\!-3~\rm {meV}$ in this section. In this case, each layer has inverted bands and contributes a Kramers pair of helical gapless edge states as shown in Fig.~\ref{fig2}(a).

By turning on the electric bias $V$, an electron Fermi surface and a hole Fermi surface are created on layer $1$ and layer $2$, respectively.
Coherent exciton condensation can be induced by the interlayer Coulomb interaction.
Throughout this paper, we focus on the time-reversal invariant $s$-wave exciton pairings which should be the leading order from the mean-field decomposition of the screened interlayer Coulomb interaction.
In general, the $s$-wave excitonic order parameters could provide an energy gain to the system,
which can be expressed as $H_{\text{X}}=\Delta_X\tau_i\sigma_js_k$ with $\Delta_X>0$ the pairing strength~(The sign change in $\Delta_X$ doesn't affect the following results) and the subscripts $i, j, k=0, x, y, z$.
The excitonic order parameters are considered to be momentum independent thanks to the short-range interaction.
There are four relevant interlayer excitonic order parameters preserving time-reversal symmetry, which are proportional to $\tau_x\sigma_zs_0, \tau_y\sigma_zs_z, \tau_y\sigma_xs_x$, and $\tau_y\sigma_xs_y$~\cite{Budich2014}. We present more details of these four excitonic order parameters in Ref.~\cite{Supplement}. 
Among these order parameters, it was found that the $\tau_x\sigma_zs_0$-type and $\tau_y\sigma_zs_z$-type orders can open a topological energy gap in the semimetallic bilayer HgTe/CdTe quantum wells, i.e., $M=0$, resulting in a helical topological excitonic insulator characterized by the $\mathbb{Z}_2$ topological invariant, while the $\tau_y\sigma_xs_x$-type and $\tau_y\sigma_xs_y$-type pairings only lead to a topologically trivial energy gap in the case of $M=0$~\cite{Budich2014}.

Next, we consider the case of $M<0$ with the $\tau_y\sigma_xs_x$-type interlayer excitonic order.
The $\tau_y\sigma_xs_y$-type order will give rise to the similar results, which is not discussed here.
In the presence of the $\tau_y\sigma_xs_x$-type order, the two pairs of helical edge states are not stable and gapped out as depicted in Fig.~\ref{fig2}(b).
When an in-plane Zeeman field $h_x$ is applied along the $x$-direction, we can see that the quasiparticle edge gap along the $k_x$-direction closes at the critical field $h_x^c$ as shown in Fig.~\ref{fig2}(c) and reopens as $h_x$ increases [see Fig. \ref{fig2}(d)].
Whereas, during this process, we verified that the edge gap along the $k_y$-direction doesn't show the closing-and-reopening behavior but only has a slight change in its amplitude.
When $h_x>h_x^c$, two distinct types of edge gaps are formed along the $x$ and $y$ directions, then we calculate the energy spectrum for a finite-sized square sample as shown in Fig.~\ref{fig2}(e). We observe four zero-energy boundary-obstructed midgap states, which are located at the four corners of the square sample by measuring the probability density~[as shown in Fig.~\ref{fig2}(f)].

\begin{figure}[hptb]
	\includegraphics[width=8cm]{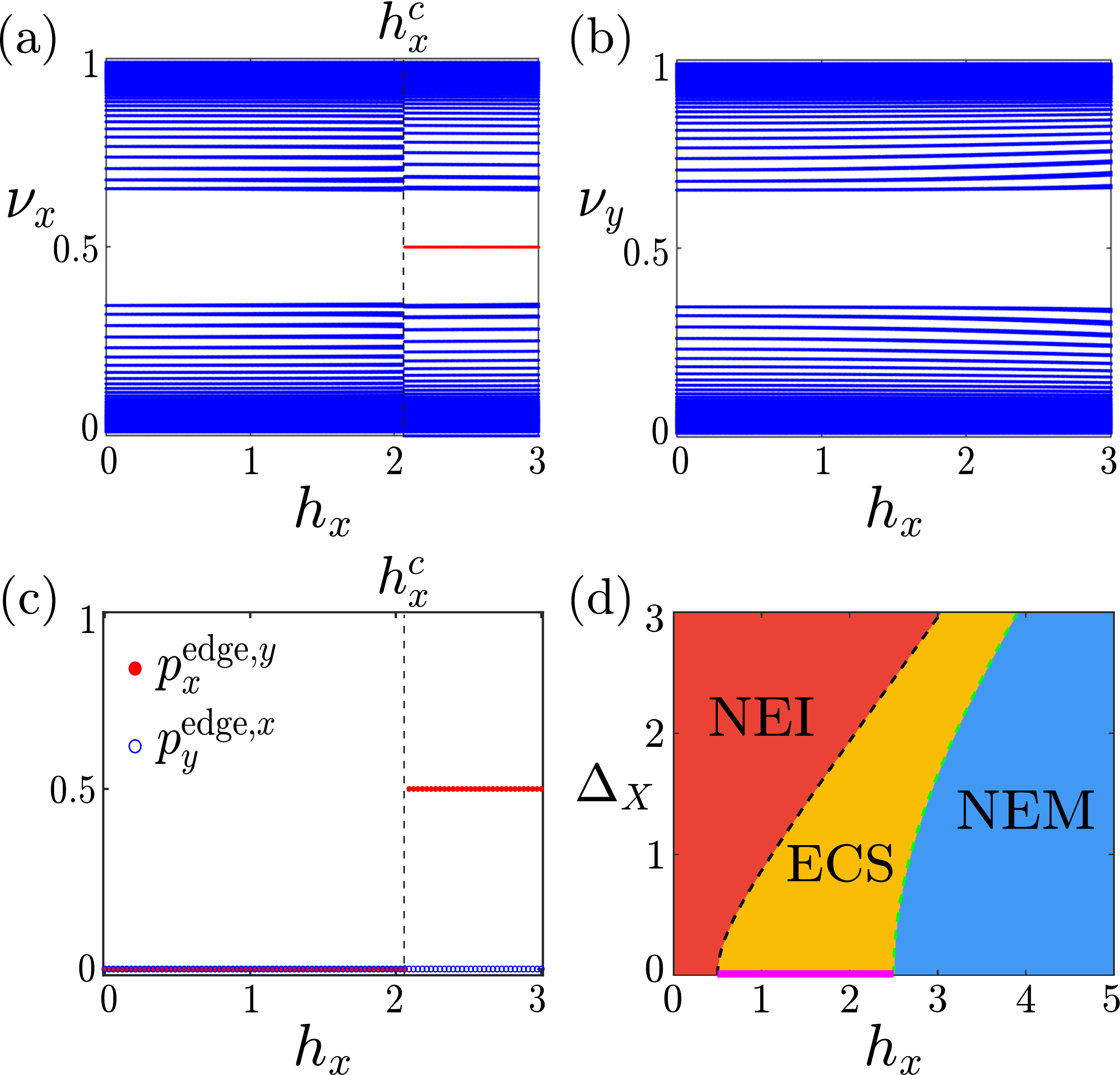} \caption{(a), (b) Wannier spectra $\nu_{x}$  and $\nu_{y}$ versus $h_{x}$. (c) Edge polarization $p_{x}^{\text{edge},y}$ and $p_{y}^{\text{edge},x}$ along $y$-normal and $x$-normal edges, respectively. In (a)-(c), $V\!=\!1~\rm {meV}$ and $\Delta_{X}\!=\!2~\rm {meV}$ are used. (d) Phase diagram for topological ECs in the $\tau_y\sigma_xs_x$-type exciton condensation on the plane formed by $\Delta_X$ and $h_x$. The red color region marked by NEI stands for the normal excitonic insulator, the orange region marked by ECS denotes the region that supports topological ECs, and the blue region means a normal excitonic metal. For $\Delta_X=0$ and $h_x>\left|V\right|/2$, the edge states are gapped along the $x$-direction but remain gapless along the $y$-direction, which marked by the solid magenta line. The dashed lines are phase boundaries determined by the topological condition for ECs $\frac{1}{2}\sqrt{\left|V\right|^2+4\Delta_X^2}\!<\! h_x \!<\! \frac{1}{2}\sqrt{(2M+\left|V\right|)^2+4\Delta_X^2}$~(See \cite{Supplement} for the details of the derivations of phase boundary conditions), which agree well with the numerical results. }%
	\label{fig3}
\end{figure}

Next, we discuss the above observed midgap states are actually the topologically protected ECs.
To unveil their topological property, we calculate the edge polarization by using the Wilson loop operators~\cite{Benalcazar2017Science,Benalcazar2017PRB,Franca2018PRB}.
For a ribbon geometry with $N_y$ unit cells in the $y$-direction and $N_{\text{orb}}$ degrees of freedom per unit cell, we express the Wilson loop operator $W_{x,k_x}$ on a path along the $k_x$-direction as $W_{x,k_x}\!=\!F_{x,k_x+(N_x-1)\Delta k_x}\!\cdot\!\cdot\!\cdot\!F_{x,k_x+\Delta k_x}F_{x,k_x}$, where $[F_{x,k_x}]^{mn}\!=\!\left\langle u_{k_x+\Delta k_x}^{m}|u_{k_x}^{n}\right\rangle$ with the step $\Delta
k_x\!=\!2\pi/N_x$, and $\left\vert u_{k_{x}}^{n}\right\rangle $ denotes the occupied Bloch functions with $n=1,...,N_{\text{occ}}$. $N_{\text{occ}}=N_{\text{orb}}N_y/2$ is the number of occupied bands.
The Wilson loop operator $W_{x,k_x}$ satisfies the following eigenvalue equation
\begin{align}
	W_{x,k_x}\left\vert \nu_{x,k_x}^j\right\rangle=e^{i2\pi\nu_x^j}\left\vert \nu_{x,k_x}^j\right\rangle,
\end{align}
where $j=1,...N_{\text{occ}}$.
We can define the Wannier Hamiltonian $H_{W_x}(k_x)$ as $W_{x,k_x}\equiv e^{iH_{W_x}(k_x)}$, of which the eigenvalues $2\pi \nu_x$ correspond to the Wannier spectrum. The tangential polarization as a function of $R_y$ is given as~\cite{Benalcazar2017Science,Benalcazar2017PRB,Franca2018PRB}
\begin{align}
	p_x(R_y)=\sum_{j=1}^{N_{\text{occ}}\times
		N_y}\rho^j(R_y)\nu_x^j,
\end{align}
where $	\rho^j(R_y)=\frac{1}{N_x}\sum_{k_x,\alpha,n}|[u_{k_x}^n]^{R_y,\alpha}[\nu_{k_x}^j]^n|^2 $ is the probability density. $[u_{k_x}^n]^{R_y,\alpha}$ with $\alpha=1,...,N_{\text{orb}}$, $R_y=1,...,N_y$ represents the components of the occupied states, and $[\nu_{x,k_x}^j]^n$ is the $n$-th component of $|\nu_{x,k_x}^j\rangle$. The edge polarization at the $y$-normal edge is defined by $p_x^{\text{edge},y}=\sum_{R_y=1}^{N_y/2}p_x(R_y)$.
To fix the sign of the polarization, we add a perturbation term $\delta\tau_y\sigma_xs_y$ in our calculations. Similarly, we can derive $\nu_y$ and $p_y^{\text{edge},x}$. We plot the Wannier spectra $\nu_x$ and $\nu_y$ as a function of $h_x$ in Figs. \ref{fig3}(a) and \ref{fig3}(b), respectively.
When the in-plane Zeeman field is greater than the critical field $h^c_x$, the ECs appear.
Correspondingly, the Wannier spectrum $\nu_x$ has a pair of values pinned at $1/2$, resulting in half quantized edge polarization $p_x^{\text{edge},y}$ shown in Fig. \ref{fig3}(c).
In contrast, $p_y^{\text{edge},x}$ remains vanishing even for $h_x>h^c_x$.
It indicates that the ECs originate from the quantized edge polarization $p_x^{\text{edge},y}$.

\emph{\color{magenta}Edge theory.}---
To provide an intuitive picture to the appearance of ECs,
we construct the edge theory~\cite{YanPRL2018} to analyze the topological mass on each edge.
For simplicity, we focus on the $V=0$ case since the bias has no contribution to the formation of edge mass.
The low-energy Hamiltonian of the exciton condensate around the $\Gamma$ point reads
\begin{align}
\begin{split}
H(\mathbf{k}) &=A\large(k_{x}\sigma_{x}s_{z}+k_{y}\sigma_{y}\large)+\left[M-B\left(k_{x}^{2}+k_{y}^{2}\right) \right]\sigma_{z}\\
	  &+\Delta_X\tau_{y}\sigma_{x}s_{x}+h_{x}s_{x}.%
\end{split}
\end{align}
We firstly consider a semi-infinite geometry occupying the space $x\geq0$ for edge I as marked in Fig. \ref{fig2}(f).
In the spirit of $\mathbf{k}\cdot\mathbf{p}$ theory, we replace $k_x\to-i\partial_x$ and divide the Hamiltonian into $H=H_0(-i\partial_x)+H_p(k_y)$, in which
\begin{align}
\begin{split}
&H_0(-i\partial_x)=-iA\sigma_{x}s_z\partial_{x}+(M+B\partial_{x}^{2})\sigma_{z},\\
&H_p(k_y)=Ak_y\sigma_{y}+\Delta_X\tau_{y}\sigma_{x}s_{x},
\end{split}
\end{align}
where all the $k_{y}^{2}$-terms have been omitted, and $h_x=0$ for edge I.
So we can solve $H_0$ first, and regard $H_p$ as a perturbation, which is justified when the exciton gap is small comparing to the energy gap.
The eigenvalue equation $H_{0}\psi_{\alpha}(x)=E_{\alpha}\psi_{\alpha}(x)$ can be solved under the boundary condition $\psi_{\alpha}(0)\!=\!\psi_{\alpha}(+\infty)\!=\!0$. A straightforward calculation gives four degenerate solutions with $E_{\alpha}=0$, whose eigenstates can be written in the following form
\begin{align}
\psi_{\alpha}(x)=N_{x}\sin(\kappa_{1}x)e^{-\kappa_{2}x}e^{ik_{y}y}\chi_{\alpha},
\end{align}
where $\alpha\!\!=\!\!1,...4$, and the normalization constant $N_{x}\!\!=\!\!2\sqrt{\kappa_{2}(\kappa_{1}^{2}+\kappa_{2}^{2})/\kappa_{1}^{2}}$ with $\kappa_{1}\!\!=\!\!\sqrt{(4BM-A^{2})/4B^{2}}$ and $\kappa_{2}\!=\!-A/2B$. The eigenvectors $\chi_{\alpha}$ are determined by $\sigma_{y}s_{z}\chi_{\alpha}=-\chi_{\alpha}$. Here we choose
\begin{align}
\begin{split}
&\chi_{1}=\left\vert \sigma_{y}=-1\right\rangle\otimes\left\vert \uparrow\right\rangle\otimes\left\vert \tau_{z}=+1\right\rangle,\\
&\chi_{2}=\left\vert \sigma_{y}=+1\right\rangle\otimes\left\vert \downarrow\right\rangle\otimes\left\vert \tau_{z}=+1\right\rangle,\\
&\chi_{3}=\left\vert \sigma_{y}=-1\right\rangle\otimes\left\vert \uparrow\right\rangle\otimes\left\vert \tau_{z}=-1\right\rangle,\\
&\chi_{4}=\left\vert \sigma_{y}=+1\right\rangle\otimes\left\vert \downarrow\right\rangle\otimes\left\vert \tau_{z}=-1\right\rangle.
\end{split}
\end{align}
In this basis set, the matrix elements of the perturbation $H_{p}(k_y)$ are represented as
\begin{align}
H_{\rm{I},\alpha\beta}(k_y)=\int_{0}^{+\infty}dx\psi_{\alpha}^{\dag}(x)H_{p}(k_y)\psi_{\beta}(x),
\end{align}
which can be written in a more compact form
\begin{align}
H_{\rm{I}}=-Ak_{y}s_{z}-\Delta_{X}\tau_{y}s_{y}.
\end{align}
Similarly, for edges II, III and IV, we obtain
\begin{align}
	\begin{split}
	H_{\rm{II}}&=-Ak_{x}s_{z}-\Delta_{X}\tau_{y}s_x+h_{x}s_{x},\\
	H_{\rm{III}}&=Ak_{y}s_{z}-\Delta_{X}\tau_{y}s_{y},\\
	H_{\rm{IV}}&=Ak_{x}s_{z}-\Delta_{X}\tau_{y}s_x+h_{x}s_{x}.
\end{split}
\end{align}
To be more clear, we introduce a unitary transformation $U=\frac{1}{\sqrt{2}}\left(
\begin{array}
	[c]{cc}%
	i & -i\\
	1 & 1%
\end{array}
\right)  $, then the edge Hamiltonians become
\begin{align}
	\begin{split}
	\widetilde{H}_{\rm{I}}&=-Ak_{y}s_{z}+\Delta_{X}\tau_{z}s_{y},\\
	\widetilde{H}_{\rm{II}}&=-Ak_{x}s_{z}+\Delta_{X}\tau_{z}s_x+h_{x}s_{x},\\
	\widetilde{H}_{\rm{III}}&=Ak_{y}s_{z}+\Delta_{X}\tau_{z}s_{y},\\
	\widetilde{H}_{\rm{IV}}&=Ak_{x}s_{z}+\Delta_{X}\tau_{z}s_x+h_{x}s_{x}.
\end{split}
\end{align}
Now all the edge Hamiltonians are block-diagonal. The effective masses of edges I and III are $M_{\rm{I}}=M_{\rm{III}}=\Delta_{X}$, while the effective masses of edges II and IV in the two blocks are $M_{\rm{II}}=M_{\rm{IV}}=\Delta_{X}+h_{x}, \Delta_X-h_x$. Therefore,  $h_x>\Delta_X$ is the topological criteria to realize the topological ECs when $V=0$.
Consequently, the effective edge masses of two adjacent boundaries have different sign, so mass domain walls appear at the intersection of these boundaries, which results in zero-energy excitonic modes according to the Jackiw-Rebbi theory~\cite{JRPRD1976}.

In Fig.~\ref{fig3}(d), we present the phase diagram of the bilayer system with the $\tau_y\sigma_xs_x$-type exciton condensate which is subjected to the in-plane Zeeman field $h_x$.
The topological excitonic corner state phase (ECS) occupies the regime between the normal excitonic insulator phase (NEI) and the normal excitonic metal phase (NEM). Here, the phase boundaries can be determined by the gap closing of the bulk at $h_x \!=\!\frac{1}{2}\sqrt{\left|V\right|^2+4\Delta_X^2}$ and $h_x \!=\!\frac{1}{2}\sqrt{(2M+\left|V\right|)^2+4\Delta_X^2}$, which agree well with the numerical results. The topological region becomes narrower by increasing the voltage $V$. Therefore, the topological ECs, characterized by the quantized edge polarization, can be realized by tuning the gate and in-plane Zeeman field.

It is necessary to point out that the in-plane Zeeman field along the $y$-direction cannot induce topological ECs in the case of $\tau_y\sigma_xs_x$-type exciton condensate. We give a more detailed explanation in Ref. \cite{Supplement} based on the edge theory.

\begin{figure}[hptb]
	\includegraphics[width=8cm]{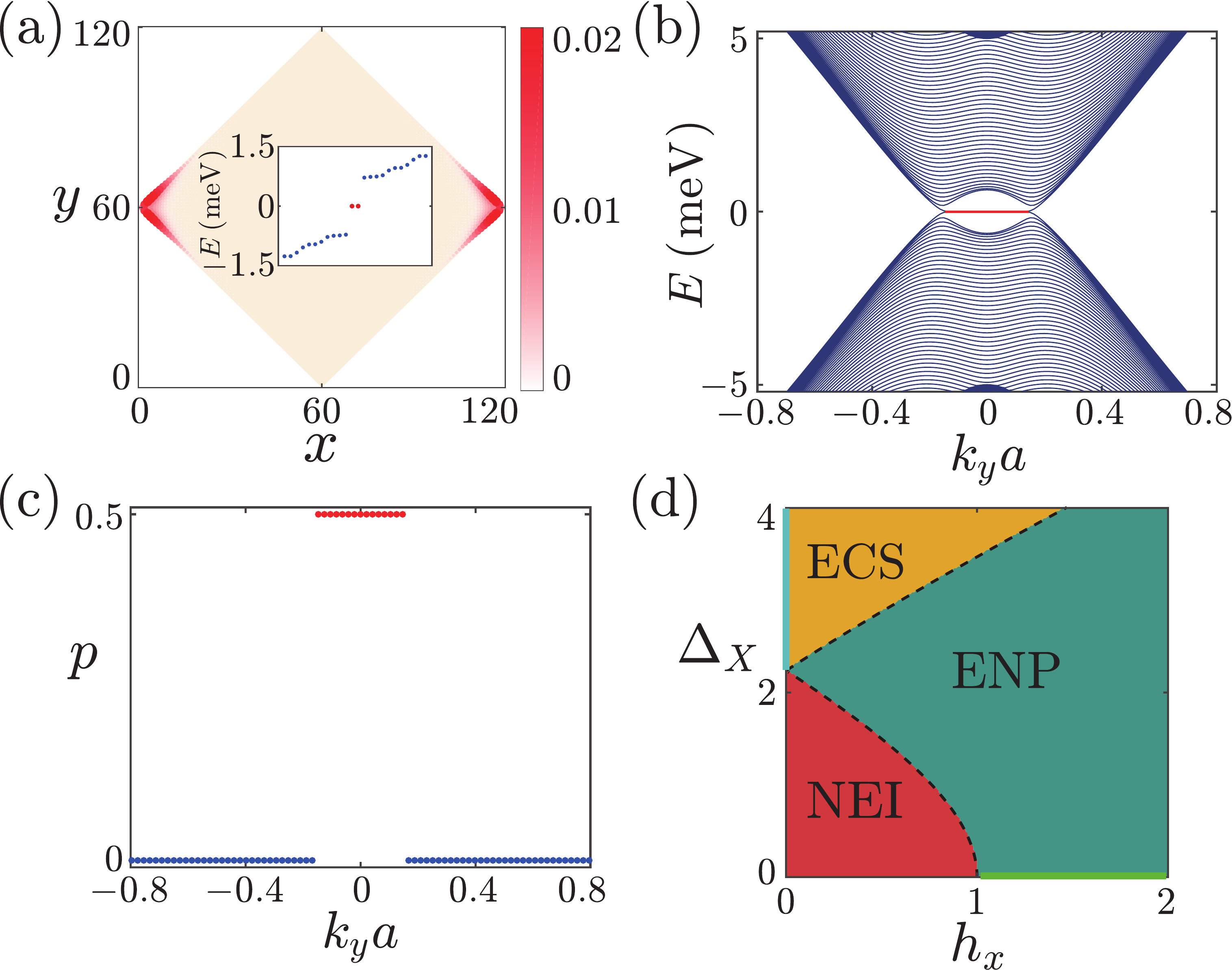} \caption{(a) Probability distribution of ECs. Inset: energy spectrum. (b) Energy dispersion of the nodal phase for a ribbon geometry with open boundary condition along the $x$-direction. (c) Bulk polarization of the nodal phase as a function of $k_y$. (d) Phase diagram of $\tau_x\sigma_zs_0$-type exciton condensation under the in-plane Zeeman field. The bottle green regime marked by ENP represents the excitonic nodal phase. Same as the Fig. \ref{fig2}(d), the red color region marked by NEI stands for the normal excitonic insulator, the orange region marked by ECS denotes the excitonic corner state phase. The solid green line means the normal metal driven by $h_x$ when $\Delta_X=0$, while the solid cyan line denotes the helical excitonic insulator induced by the $\tau_x\sigma_zs_0$-type pairing when $h_x=0$. In (a), (b) and (c), we set $V\!\!=$ 4 meV, $\Delta_{X}\!=$ 5 meV, $h_{x}\!=$ 3 meV for (b) and (c), and $h_{x}\!=$ 1 meV for (a). In all plots, we choose $M=3~\rm {meV}$.}%
	\label{fig4}
\end{figure}

\emph{\color{magenta}Excitonic corner states and nodal phase for the positive Dirac mass.}---
Next, we discuss if we can still realize the topological ECs in the bilayer system without band inversion occurring in each layer. For the $\tau_x\sigma_zs_0$-type exciton condensate, the in-plane Zeeman filed can also generate ECs even when $H_{\rm{QSH}}$ has non-negative mass $M\geq0$.
For our purpose, we set $M\!=\!3~\rm {meV}$ in this section. Note that the bilayer system is a topologically trivial semiconductor in the absence of excitonic orders.
The numerical calculation shows the in-plane Zeeman field $h_x$ induces two ECs located at the left and right corners for a diamond-shaped sample [see Fig. \ref{fig4}(a)].

Although the Zeeman field breaks time-reversal symmetry, $h_x$ preserves the 2-fold rotation symmetry about the $x$-axis $C_{2x}\!=\!i\sigma_{z}s_x$. The first Brillouin zone has a mirror-invariant line $k_{y}\!=\!0$ preserving the $C_{2x}$ symmetry. We can adopt the mirror winding number~\cite{Park2019PRL} along this line to characterize the topological properties of this type of ECs. In this line, $H(k_x,k_y=0)$ is invariant under the operation of $C_{2x}$. The discrete version of $H(k_x,0)$ has the following form
\begin{align}
	H(k_{x},0)=&\frac{A}{a} \sin(k_{x}a)\sigma_{x}s_{z}+M(k_{x},0)\sigma_{z}\nonumber\\
	+&h_{x}s_{x}-\frac{V}{2}\tau_{z}+\Delta_{X}\tau_{x} \sigma_{z},
\end{align}
where $M(k_{x},0)=M-B[2-2\cos(k_{x}a)]/a^2$. $C_{2x}$ has two fourfold degenerate eigenvalues of $\pm 1$, the eigenvectors of $\pm 1$ are $1/\sqrt{2}\left\vert \alpha\right\rangle\otimes(\left\vert \downarrow\right\rangle \pm \left\vert \uparrow\right\rangle)\otimes \left\vert l=2\right\rangle$,$1/\sqrt{2}\left\vert \beta\right\rangle\otimes(\left\vert \downarrow\right\rangle \mp \left\vert \uparrow\right\rangle)\otimes \left\vert l=2\right\rangle$, $1/\sqrt{2}\left\vert \alpha\right\rangle\otimes(\left\vert \downarrow\right\rangle \pm \left\vert \uparrow\right\rangle)\otimes \left\vert l=1\right\rangle$,$1/\sqrt{2}\left\vert \beta\right\rangle\otimes(\left\vert \downarrow\right\rangle \mp \left\vert \uparrow\right\rangle)\otimes \left\vert l=1\right\rangle$, where $\left\vert \alpha\right\rangle(\left\vert \beta\right\rangle)$, $\left\vert \uparrow\right\rangle(\left\vert \downarrow\right\rangle)$ and $\left\vert l=1,2\right\rangle(\left\vert 2\right\rangle)$ are the basis vectors acting on the orbit, spin and layer subspaces, respectively.
Due to the conserving of $C_{2x}$, we project $H(k_{x},0)$ into the two subspaces corresponding to $C_{2x}=\pm 1$, i.e., $H(k_{x},0) = H_{+}(k_x,0)  \oplus H_{-}(k_{x},0)$.
And the block Hamiltonians read
\begin{align}
\begin{split}
H_{\pm}(k_{x},0)&=[M(k_{x},0) \pm h_{x}]\sigma_{z}-\frac{V}{2}\tau_{z}\\
&-\frac{A}{a}\sin(k_{x} a)\sigma_{x}+\Delta_{X}\tau_{x}\sigma_{z}.
\end{split}
\end{align}
 Along the line $k_{y}=0$, we consider the Wilson loop operator $W_{\pm,k_{x}}$, then the mirror winding number $\nu_{\pm}$ can be evaluated by~\cite{Park2019PRL}
\begin{align}
\nu_{\pm}=\frac{1}{i\pi}\log(\det[W_{\pm,k_{x}}])~\rm{mod}~2.
\end{align}
When the ECs emerge, the mirror winding number shows that $\nu_+=\nu_-= 1$.

Now let us discuss the excitonic order induced nodal phase in the bilayer system.
In the case of the $\tau_x\sigma_zs_0$-type exciton pairing, a nodal phase with Weyl nodes along the $k_y$-axis emerges. The excitonic nodal phase hosts flat band edge states as shown in Fig. \ref{fig4}(b). In order to characterize the topological properties of nodal phase, we use the Wilson loop method to calculate the bulk polarization of the system. By treating $k_y$ as a parameter, the Hamiltonian is effectively reduced to a one-dimensional Hamiltonian $H_{k_y}(k_x)$. For fixed $k_y$, considering the Wilson loop operator in the $x$-direction
$W_{x,k_x}$, the Wannier center $\nu_{x}^{j}$ is obtained by the following equation
\begin{align}
	W_{x,k_{x}}\left\vert \nu_{x,k_x}^{j}\right\rangle=e^{i2\pi\nu_{x}^{j}}\left\vert \nu_{x,k_x}^{j}\right\rangle.
\end{align}
Then, the bulk polarization can be defined as $p=\sum_{j} \nu_{x}^{j}~\rm{mod}~1$ for a given $k_y$. In Fig. \ref{fig4}(c), we plot the calculated bulk polarization as a function of $k_y$. We can see that the polarization is quantized to $1/2$ between two nodes and vanishes at other $k_y$. Therefore, the topology of the nodal phase can be captured by the $k_y$-dependent polarization.

Finally, the phase diagram for $\tau_x\sigma_zs_0$-type exciton condensate on the plane of $\Delta_X$ and $h_x$ is shown in Fig. \ref{fig4}(d).
By numerically observing the gap closing of the bulk, we define the phase boundaries. Analytically, the boundary between the ECS and the excitonic nodal phase (ENP) is determined by $h_{x}\!=\!-M+\sqrt{\Delta_{X}^{2}+\left|V\right|^{2}/4}$, while the boundary between the NEI and ENP is defined by $h_{x}\!=\!M-\sqrt{\Delta_{X}^{2}+\left|V\right|^{2}/4}$~\cite{Supplement}.

\emph{\color{magenta}Conclusion and discussion.}---In this work, we identified two distinct types of ECs in the gated bilayer quantum spin Hall insulator model with $s$-wave exciton pairings in the presence of the in-plane Zeeman field. Experimentally, the ECs can be detected by Scanning Tunning Microscope measurements. ECs manifest themselves as in-plane Zeeman field dependent zero-bias peaks in differential conductance~(See Section V in ~\cite{Supplement} for more details). The different patterns of the two types of ECs, in turn, could be used to determine the excitonic pairing of the excitonic insulator in experiments. We also found an excitonic nodal phase with the flat-band edge states in this system. Considering these exotic topological phases, our work will stimulate more investigations on higher-order topology and topological nodal phases in exciton condensates.

Different from superconducting pairings, excitonic pairings don't have to possess particle-hole symmetry. Therefore, ECs can appear at the finite energy, which is in contrast to Majorana corner states. In this paper, the ECs are pinned to zero energy as we use a particle-hole symmetric model. Removing particle-hole symmetry, we can still expect midgap ECs, but they will be shifted to the finite energy.

Additionally, we mainly focus on the corner states created in the time-reversal invariant singlet $s$-wave exciton condensates hereinbefore. In this case, an in-plane Zeeman field is necessary to create the ECs.
Whereas, we would like to point out that Kramers pairs of ECs could be generated in this bilayer system without applying a Zeeman field when time-reversal invariant $d$-wave exciton pairings are formed.

\emph{\color{magenta}Acknowledgments.}---
D.-H.X. was supported by the NSFC (under Grant Nos. 12074108 and 11704106). B.Z. was supported by the NSFC (under Grant No. 12074107) and the program of outstanding young and middle-aged scientific and technological innovation team of colleges and universities in Hubei Province (under Grant No. T2020001). C.-Z.C. was funded by the NSFC (under Grant No. 11974256) and the NSF of Jiangsu Province (under Grant No. BK20190813). D.-H.X. also acknowledges the financial support of the Chutian Scholars Program in Hubei Province.

\bibliography{bibfile}

\pagebreak
\widetext
\clearpage

\setcounter{equation}{0}
\setcounter{figure}{0}
\makeatletter
\renewcommand{\theequation}{S\arabic{equation}}
\renewcommand{\thefigure}{S\arabic{figure}}
\renewcommand{\bibnumfmt}[1]{[S#1]}
\renewcommand{\citenumfont}[1]{S#1}

\begin{center}
\textbf{\large Supplemental Material to: ``Topological excitonic corner states and nodal phase in bilayer quantum spin Hall insulators''}
\end{center}

In this Supplemental Material, we give more details of the four excitonic order parameters in the main text used to study topological excitonic corner states. Subsequently, we show the different behaviors of the in-plane Zeeman fields along the $x$-direction and $y$-direction in producing the excitonic corner states for excitonic pairings $\Delta_{X}\tau_{y}\sigma_{x}s_{x}$ and $\Delta_{X}\tau_{y}\sigma_{x}s_{y}$. Besides, we also present the details of the derivation of the phase boundary conditions. Finally, we provide the local density of states plots, which can be used to detect the corner states in Scanning Tunning Microscope (STM) probe.

\section{Time-invariant excitonic order parameters}
\label{I}
The low-energy effective Hamiltonian of the gated bilayer quantum spin Hall insulator model in the momentum space is given by
\begin{equation}\label{HQSH}
H_{\rm QSH}(\mathbf{k})=M(\mathbf{k})\tau_{0}\sigma_{z}s_{0}+A k_{x}\tau_{0}\sigma_{x}s_{z}+A k_{y}\tau_{0}\sigma_{y}s_{0}-\frac{V}{2}\tau_{z}\sigma_{0}s_{0},
\end{equation}
where $s_{x,y,z}$, $\sigma_{x,y,z}$ and $\tau_{x,y,z}$ are Pauli matrices acting on the spin, orbital and layer degrees of freedom, respectively. $\tau_{0}$, $\sigma_{0}$ and $s_{0}$ are the $2\times2$ identity matrices. The Hamiltonian $H_{\rm QSH}(\mathbf{k})$ preserves time-reversal symmetry, $\mathcal{T}H_{\rm QSH}(\mathbf{k})\mathcal{T}^{-1}=H_{\rm QSH}(-\mathbf{k})$ with $\mathcal{T}=i\tau_{0}\sigma_{0}s_{y}\mathcal{K}$ (where $\mathcal{K}$ is the complex conjugation). Meanwhile, $H_{\rm QSH}(\mathbf{k})$ also has inversion symmetry with the inversion operator $P=\tau_z\sigma_z$. We focus on the uniform momentum-independent $s$-wave-like excitonic order parameters induced by interlayer Coulomb interaction. In general, the excitonic pairing term could be written as $H_{X}=\Delta_{X}\tau_{i}\sigma_{j}s_{k}$ with $\Delta_{X}$ the uniform pairing strength and the subscripts $i,j,k=0,x,y,z$. There are in total 16 interlayer excitonic order parameters preserving time-reversal symmetry, which are represented as
\begin{align}
\tau_{x}\sigma_{0}s_{0}, \tau_{x}\sigma_{x}s_{0}, \tau_{x}\sigma_{y}s_{x}, \tau_{x}\sigma_{y}s_{y},\nonumber\\
\tau_{x}\sigma_{y}s_{z}, \tau_{x}\sigma_{z}s_{0}, \tau_{y}\sigma_{0}s_{x}, \tau_{y}\sigma_{0}s_{y},\nonumber\\
\tau_{y}\sigma_{0}s_{z}, \tau_{y}\sigma_{x}s_{x}, \tau_{y}\sigma_{x}s_{y}, \tau_{y}\sigma_{x}s_{z},\nonumber\\
\tau_{y}\sigma_{y}s_{0}, \tau_{y}\sigma_{z}s_{x}, \tau_{y}\sigma_{z}s_{y}, \tau_{y}\sigma_{z}s_{z}.
\end{align}

Among these order parameters, $\tau_{x}\sigma_{z}s_{0}$-type and $\tau_{y}\sigma_{z}s_{z}$-type commute with the mass term and anticommute with the second and third terms in Eq.~(\ref{HQSH}), hence they renormalize the Dirac mass. These two order parameters can lead to the time-invariant topological excitonic insulator with helical edge states. Note that, these two order parameters break inversion symmetry. The other relevant order parameters are $\tau_{y}\sigma_{x}s_{x}$-type and $\tau_{y}\sigma_{x}s_{y}$-type, which anticommute with $H_{\rm QSH}(\mathbf{k})$. In the case of negative Dirac mass, the helical edge states of the quantum spin Hall insulator couple together to open an energy gap in the presence of $\tau_{y}\sigma_{x}s_{x}$-type and $\tau_{y}\sigma_{x}s_{y}$-type order parameters. Applying in-plane Zeeman fields, these two classes of excitonic order parameters, including $\tau_x\sigma_zs_0, \tau_y\sigma_zs_z, \tau_y\sigma_xs_x$, and $\tau_y\sigma_xs_y$, can give rise to topological excitonic corner states. Therefore, we focus on these two classes excitonic order parameters in the main text.


\section{Comments on the in-plane Zeeman field}
\label{II}
In this Section, by using the effective edge theory, we demonstrate the different behaviors of the in-plane Zeeman fields along the $x$ and $y$ directions in producing corner states for the excitonic pairings $\Delta_{X}\tau_{y}\sigma_{x}s_{x}$ and $\Delta_{X}\tau_{y}\sigma_{x}s_{y}$.


For our purpose, we consider an in-plane Zeeman field with both $x$ and $y$ components. Then we construct the edge theory to analyze the effective mass on each edge. For simplicity, we ignore the bias term as it doesn't contribute to the effective edge mass. First, let us consider the case of $\Delta_{X}\tau_{y}\sigma_{x}s_{x}$-type pairing, then low-energy Hamiltonian around the $\Gamma$ point reads
\begin{align}
H(\mathbf{k})=A(k_{x}\sigma_{x}s_{z}+k_{y}\sigma_{y})+[M-B(k_{x}^{2}+k_{y}^{2})]\sigma_{z}+\Delta_{X}\tau_{y}\sigma_{x}s_{x}
+h_{x}s_{x}+h_{y}s_{y}.
\end{align}
We firstly consider a semi-infinite geometry occupying the space $x\ge 0$ for edge I as marked in Fig. 1(f) of the main text. In the spirit of $\textbf{k}\cdot\textbf{p}$ theory, we replace $k_{x}\rightarrow -i\partial_{x}$ and separate the Hamiltonian into $H=H_{0}(-i\partial_{x})+H_{p}(k_{y})$, in which
\begin{align}
&H_{0}(-i\partial_{x})=-i A\sigma_{x}s_{z}\partial_{x}+(M+B\partial_{x}^{2})\sigma_{z},\nonumber\\
&H_{p}(k_{y})=A k_{y}\sigma_{y}+\Delta_{X}\tau_{y}\sigma_{x}s_{x}+h_{x}s_{x}+h_{y}s_{y},
\end{align}
where all the $k_{y}^{2}$-terms have been neglected. Hence we can solve $H_{0}$ first, and treat $H_{p}$ as a perturbation. The eigenvalue equation $H_{0}\psi_{\alpha}(x)=E_{\alpha}\psi_{\alpha}(x)$ can be solved under the boundary condition $\psi_{\alpha}(0)=\psi_{\alpha}(+\infty)=0$. A straightforward calculation gives four degenerate solutions with $E_{\alpha}=0$, whose eigenstates can be written in the following form
\begin{align}
\psi_{\alpha}(x)=N_{x}\sin(\kappa_{1}x)e^{-\kappa_{2}x}e^{i k_{y}y}\chi_{\alpha},
\end{align}
where $\alpha=1,...4$, and the normalization constant $N_{x}=2\sqrt{\kappa_{2}(\kappa_{1}^{2}+\kappa_{2}^{2})/\kappa_{1}^{2}}$ with $\kappa_{1}=\sqrt{(4BM-A^{2})/4 B^{2}}$ and $\kappa_{2}=-A/2B$. The eigenvectors $\chi_{\alpha}$ are determined by $\sigma_{y}s_{z}\chi_{\alpha}=-\chi_{\alpha}$. Here we choose
\begin{align}
\chi_{1}=\left\vert \sigma_{y}=-1 \right\rangle \otimes \left\vert \uparrow \right\rangle \otimes \left\vert \tau_{z}=+1 \right\rangle,\nonumber\\
\chi_{2}=\left\vert \sigma_{y}=+1 \right\rangle \otimes \left\vert \downarrow \right\rangle \otimes \left\vert \tau_{z}=+1 \right\rangle,\nonumber\\
\chi_{3}=\left\vert \sigma_{y}=-1 \right\rangle \otimes \left\vert \uparrow \right\rangle \otimes \left\vert \tau_{z}=-1 \right\rangle,\\
\chi_{4}=\left\vert \sigma_{y}=+1 \right\rangle \otimes \left\vert \downarrow \right\rangle \otimes \left\vert \tau_{z}=-1 \right\rangle.\nonumber
\end{align}
In this basis set, the matrix elements of the perturbation $H_{p}(k_{y})$ are represented as
\begin{align}
H_{\text{I},\alpha\beta}(k_{y})=\int_{0}^{+\infty}dx\psi_{\alpha}^{\dagger}(x)H_{p}(k_{y})\psi_{\beta}(x),
\end{align}
which can be written in a more compact form
\begin{align}
H_{\text{I}}=-A k_{y}\tau_{0}s_{z}-\Delta_{X}\tau_{y}s_{y}.
\end{align}
For edge II, the separated Hamiltonians are
\begin{align}
&H_{0}\left(  -i\partial_{y}\right)  =-iA\tau_{0}\sigma_{y}s_{0}\partial_{y}+\left(  M+B\partial_{y}^{2}\right)  \tau_{0}\sigma_{z}s_{0},\nonumber\\
&H_{p}\left(  k_{x}\right)  =Ak_{x}\tau_{0}\sigma_{x}s_{z}+\Delta_{X}\tau
_{y}\sigma_{x}s_{x}+h_{x}s_{x}+h_{y}s_{y}.
\end{align}
We choose the basis
\begin{align}
	\chi_{1}  &  =\left\vert \sigma_{x}=-1\right\rangle \otimes\left\vert
	\uparrow\right\rangle \otimes\left\vert \tau_{z}=+1\right\rangle ,\nonumber\\
	\chi_{2}  &  =\left\vert \sigma_{x}=-1\right\rangle \otimes\left\vert
	\downarrow\right\rangle \otimes\left\vert \tau_{z}=+1\right\rangle ,\nonumber\\
	\chi_{3}  &  =\left\vert \sigma_{x}=-1\right\rangle \otimes\left\vert
	\uparrow\right\rangle \otimes\left\vert \tau_{z}=-1\right\rangle ,\nonumber\\
	\chi_{4}  &  =\left\vert \sigma_{x}=-1\right\rangle \otimes\left\vert
	\downarrow\right\rangle \otimes\left\vert \tau_{z}=-1\right\rangle,
\end{align}
which satisfies $\tau_{0}\sigma_{x}s_{0}\xi_{\alpha}=-\xi_{\alpha}$. In this basis, we have
\begin{align}
	H_{\text{II}}=-A k_{x}\tau_{0}s_{z}-\Delta_{X}\tau_{y}s_{x}+h_{x}\tau_{0}s_{x}+h_{y}\tau_{0}s_{y}.
\end{align}
Similarly, for edges III and IV, we obtain
\begin{align}
&H_{\text{III}}=A k_{y}\tau_{0}s_{z}-\Delta_{X}\tau_{y}s_{y},\\
&H_{\text{IV}}=A k_{x}\tau_{0}s_{z}-\Delta_{X}\tau_{y}s_{x}+h_{x}\tau_{0}s_{x}+h_{y}\tau_{0}s_{y}.\nonumber
\end{align}
When $\Delta_{X}=0$, we can see that both the $x$-component and $y$-component of the in-plane Zeeman field can only open an energy gap on edges II and IV, while the edge states on edges I and III are not affected by the in-plane Zeeman field. In contrast, the excitonic pairing $\Delta_{X}\tau_{y}\sigma_{x}s_{x}$ produces a uniform gap for edge states.

To be more clear, we introduce a unitary transformation $U=\frac{1}{\sqrt{2}}\left( \begin{array}{cc}
i & -i \\
1 & 1%
\end{array}\right)$, then the edge Hamiltonians become
\begin{align}
&\widetilde{H}_{\text{I}}=-A k_{y}\tau_{0}s_{z}+\Delta_{X}\tau_{z}s_{y},\nonumber\\
&\widetilde{H}_{\text{II}}=-A k_{x}\tau_{0}s_{z}+\Delta_{X}\tau_{z}s_{x}+h_{x}\tau_{0}s_{x}+h_{y}\tau_{0}s_{y},\nonumber\\
&\widetilde{H}_{\text{III}}=A k_{y}\tau_{0}s_{z}+\Delta_{X}\tau_{z}s_{y},\\
&\widetilde{H}_{\text{IV}}=A k_{x}\tau_{0}s_{z}+\Delta_{X}\tau_{z}s_{x}+h_{x}\tau_{0}s_{x}+h_{y}\tau_{0}s_{y}.\nonumber
\end{align}
The eigenvalues are obtained by the diagonalizing the edge Hamiltonians

\begin{align}
&E_{\text{I}},E_{\text{III}}=-\sqrt{A^{2}k_{y}^{2}+\Delta_{X}^{2}},\sqrt{A^{2}k_{y}^{2}+\Delta_{X}^{2}},\nonumber\\
&E_{\text{II}},E_{\text{IV}}=-\sqrt{A^{2}k_{x}^{2}+(\Delta_{X}+h_{x})^{2}+h_{y}^{2}},
\sqrt{A^{2}k_{x}^{2}+(\Delta_{X}+h_{x})^{2}+h_{y}^{2}},\nonumber\\
&-\sqrt{A^{2}k_{x}^{2}+(\Delta_{X}-h_{x})^{2}+h_{y}^{2}},
\sqrt{A^{2}k_{x}^{2}+(\Delta_{X}-h_{x})^{2}+h_{y}^{2}}.
\end{align}
Apparently, for the $\tau_{y}\sigma_{x}s_{x}$-type pairing, only the $h_x$-component of the in-plane Zeeman field can drive a phase transition on edges II and IV in which the energy gap opened by the excitonic paring closes and reopens by increasing $h_x$. As stated in the main text, this phase transition is the key to realize the excitonic corner states. In the case of $\tau_{y}\sigma_{x}s_{y}$-type excitonic pairing, however, $h_y$ is responsible for the closing and reopening of the energy gap on edges II and IV.


%

\section{Phase transition point in the case of $\tau_{y}\sigma_{x}s_{x}$-type pairing}
\label{III}
In this section, we give the specific derivation process of the analytical expression of the phase transition points in Fig. 3(d). Here we first discuss the phase transition in the case of $\tau_{y}\sigma_{x}s_{x}$-type pairing. The Hamiltonian $\mathcal{H}(\mathbf{k})$ of k-space is,
\begin{align}
\mathcal{H}(\textbf{k})=&[M-B(k_{x}^{2}+k_{y}^{2})]\tau_{0}\sigma_{z}s_{0}+A(k_{x}\tau_{0}\sigma_{x}s_{z}+k_{y}\tau_{0}\sigma_{y}s_{0})\nonumber\\
&-\frac{V}{2}\tau_{z}+\Delta_{X}\tau_{y}\sigma_{x}s_{x}+h_{x}\tau_{0}\sigma_{0}s_{x},
\end{align}
where the basis is $c_{\textbf{k}}^{\dagger}=(c_{\textbf{k}1\alpha\uparrow}^{\dagger},c_{\textbf{k}1\alpha\downarrow}^{\dagger},
c_{\textbf{k}1\beta\uparrow}^{\dagger},c_{\textbf{k}1\beta\downarrow}^{\dagger},c_{\textbf{k}2\alpha\uparrow}^{\dagger},
c_{\textbf{k}2\alpha\downarrow}^{\dagger},c_{\textbf{k}2\beta\uparrow}^{\dagger},c_{\textbf{k}2\beta\downarrow}^{\dagger})$, $\alpha$ and $\beta$ are different orbital degrees of freedom with opposite parity, $\uparrow$ and $\downarrow$ represent electron spin, and $1,2$ are the layer indices.
Since the energy gap of the bilayer HgTe quantum wells is at the $\Gamma (k_{x}\!=\!0,k_{y}\!=\!0)$ point, we diagonalize  $\mathcal{H}(\textbf{k})$ and obtain eight eigenvalues at $\Gamma$
\begin{align}
\mathcal{E}=\pm h_{x} \pm \frac{1}{2}\sqrt{(2M \pm V)^{2}+4\Delta_{X}^{2}},
\end{align}
In the case of $\tau_{y}\sigma_{x}s_{x}$-type pairing, we set $M\!<\!0$, therefore, when $h_{x}=\frac{1}{2}\sqrt{(2M+|V|)^{2}+4\Delta_{X}^{2}}$, the bulk energy gap is closed, and when $h_{x}>\frac{1}{2}\sqrt{(2M+|V|)^{2}+4\Delta_{X}^{2}}$, the system exhibits a metallic state. Meanwhile, the excitonic corner states are formed by gapping out the helical edge states, thus we can used the effective Hamiltonian obtained by the edge theory to determine the regime of the corner states. The effective Hamiltonians with bias are as follows:
\begin{align}
&\widetilde{\mathcal{H}}_{\text{I}}=-A k_{y}\tau_{0}s_{z}+\frac{V}{2}\tau_{x}s_{0}+\Delta_{X}\tau_{z}s_{y},\nonumber\\
&\widetilde{\mathcal{H}}_{\text{II}}=-A k_{x}\tau_{0}s_{z}+\frac{V}{2}\tau_{x}s_{0}+\Delta_{X}\tau_{z}s_{x}+h_{x}\tau_{0}s_{x},\nonumber\\
&\widetilde{\mathcal{H}}_{\text{III}}=A k_{y}\tau_{0}s_{z}-\frac{V}{2}\tau_{x}s_{0}+\Delta_{X}\tau_{z}s_{y},\\
&\widetilde{\mathcal{H}}_{\text{IV}}=A k_{x}\tau_{0}s_{z}-\frac{V}{2}\tau_{x}s_{0}+\Delta_{X}\tau_{z}s_{x}+h_{x}\tau_{0}s_{x}.\nonumber
\end{align}
The eigenvalues are obtained by the diagonalizing the edge Hamiltonians
\begin{align}
&\mathcal{E}_{\text{I}}, \mathcal{E}_{\text{III}}=\pm \frac{1}{2}\sqrt{(2A k_{y} \pm V)^{2}+4\Delta_{X}^{2}},\nonumber\\
&\mathcal{E}_{\text{II}}, \mathcal{E}_{\text{IV}}=\pm \frac{1}{2}\sqrt{4\Delta_{X}^{2}+V^{2} \pm 4\sqrt{A^{2}V^{2}k_{x}^{2}+(V^{2}+4\Delta_{X}^{2})h_{x}^{2}}+4A^{2}k_{x}^{2}+4h_{x}^{2}}.
\end{align}
When $k_{x}=0$ and $k_{y}=0$,
\begin{align}
&\mathcal{E}_{\text{I}}, \mathcal{E}_{\text{III}}=\pm \frac{1}{2}\sqrt{V^{2}+4\Delta_{X}^{2}},\nonumber\\
&\mathcal{E}_{\text{II}}, \mathcal{E}_{\text{IV}}=\pm \sqrt{(\sqrt{\frac{V^{2}}{4}+\Delta_{X}^{2}} \pm h_{x})^{2}}.
\end{align}
Obviously, when $V,\Delta_{X} \ne 0$, the energy gap between edge I and III will always exist. However, when $h_{x}=\frac{1}{2}\sqrt{V^{2}+4\Delta_{X}^{2}}$, the energy gap of edge II and IV will close, and the energy gap will reopen and corner states emerge when $h_{x}$ continues to increase.

In summary, when $h_{x} \!<\! \frac{1}{2}\sqrt{V^{2}+4\Delta_{X}^{2}}$, the edge states of quantum spin Hall insulators are gapped by excitonic pairing $\Delta_{X}$. At this time, the system is in the normal excitonic insulator (NEI) phase. When $h_{x}\!>\!\frac{1}{2}\sqrt{(2M+|V|)^{2}+4\Delta_{X}^{2}}$, the bulk energy gap is closed and the system behaves as normal excitonic metal (NEM). Only when $\frac{1}{2}\sqrt{V^{2}+4\Delta_{X}^{2}}<\! h_{x}\!<\frac{1}{2}\sqrt{(2M+|V|)^{2}+4\Delta_{X}^{2}}$, we can observe excitonic corner states(ECS).

\section{Phase transition point in the case of $\tau_{x}\sigma_{z}s_{0}$-type pairing}
\label{IV}
In this section, we give the specific derivation process of the analytical expression of the phase boundary of Fig. 4(d) in the main text.
Here we consider the case where the excitonic pairing of the system is $\tau_{x}\sigma_{z}s_{0}$-type, and we have $M>0$. The Hamiltonian of the system
\begin{align}
H(\textbf{k})=&[M-B(k_{x}^{2}+k_{y}^{2})]\tau_{0}\sigma_{z}s_{0}+A(k_{x}\tau_{0}\sigma_{x}s_{z}+k_{y}\tau_{0}\sigma_{y}s_{0})\nonumber\\
&-\frac{V}{2}\tau_{z}+\Delta_{X}\tau_{x}\sigma_{z}s_{0}+h_{x}\tau_{0}\sigma_{0}s_{x},
\end{align}
By diagonalization, the eigenvalues of the Hamiltonian at $\Gamma$ point are given by
\begin{align}
E=\pm h_{x} \pm M \pm \frac{1}{2}\sqrt{V^{2}+4\Delta_{X}^{2}},
\end{align}
Obviously, when $h_{x}=-M+\frac{1}{2}\sqrt{V^{2}+4\Delta_{X}^{2}}$ or $h_{x}=M-\frac{1}{2}\sqrt{V^{2}+4\Delta_{X}^{2}}$, the bulk energy gap of the system will be closed. This is exactly the two dashed lines denoting the phase boundary marked in Fig. 4(d) of the main text. In addition, we can confirm the regime between the two dashed lines are excitonic corner states through the mirror winding number.

\section{Local density of states and STM detection}
\label{V}

\begin{figure}[hptb]
	\includegraphics[width=14cm]{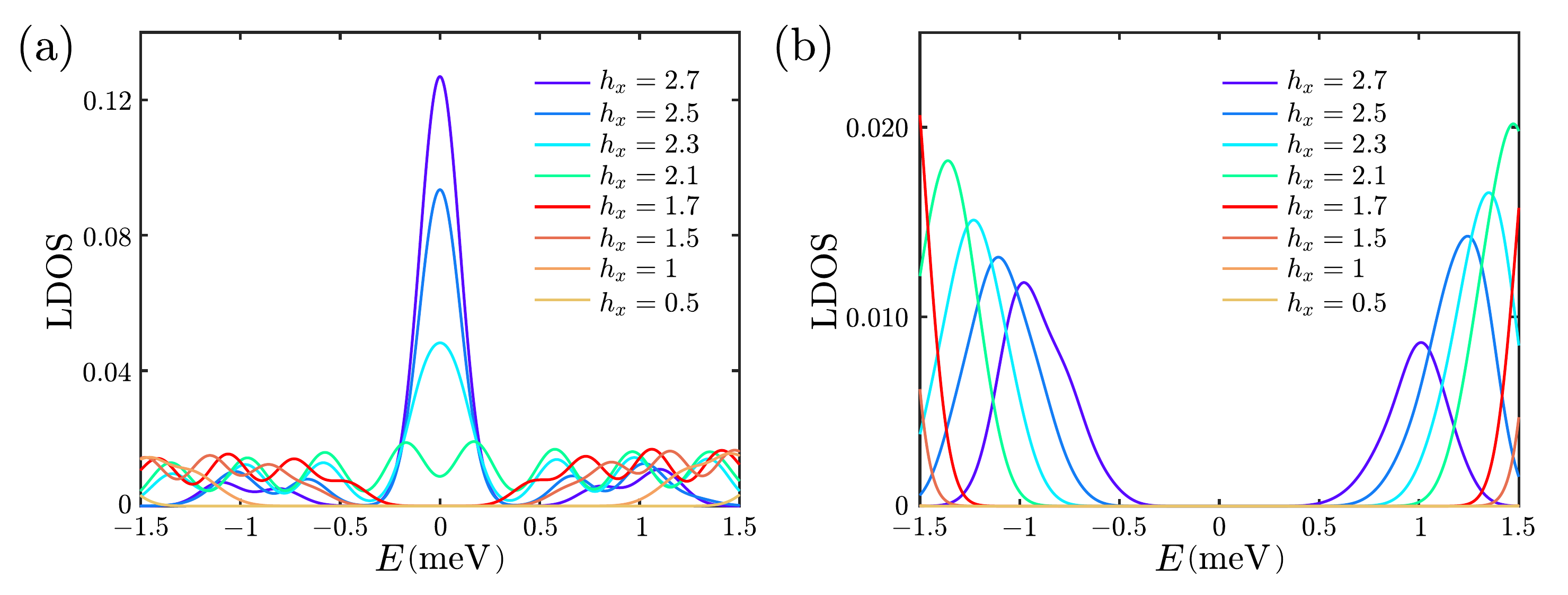} \caption{(a) The local density of states at the corner \textbf{x} ($x\!=\!1, y\!=\!1$) for different in-plane Zeeman field strength. (b) The local density of states at the edge \textbf{x} ($x\!=\!1, y\!=\!L_{y}/2$) for different in-plane Zeeman field strength. In all plots, we choose $M=-3$ meV, $V=1$ meV, $\Delta_{X}=2$ meV. The size of the square-shaped sample is $L_{x}\!\times\!L_{y}=100\!\times\!100$.}%
	\label{figS1}
\end{figure}

Experimentally, we predict that Scanning Tunning Microscope (STM) probes can be used to detect the existence of corner modes. In this section, we focus on the experimental suggestion on the excitonic corner states in the case of $\tau_{y}\sigma_{x}s_{x}$-type pairing. 
We know that STM serves as a probe of the local density of states (LDOS) of the sample. We assume that the STM probe detects layer 1 of the system (our calculations show that there is no essential difference between detecting layer 1 and layer 2).
The local density of states is defined as $\rho (E, l\!=\!1, \textbf{x})=\sum_{i}\delta (E-E_{i})(|\Psi_{E_{i},1\alpha\uparrow\textbf{x}}|^{2}+|\Psi_{E_{i},1\alpha\downarrow\textbf{x}}|^{2}
+|\Psi_{E_{i},1\beta\uparrow\textbf{x}}|^{2}+|\Psi_{E_{i},1\beta\downarrow\textbf{x}}|^{2})$, where $E_{i}$ is the $i$th eigenvalue,  $l=1,2$, $\sigma=\alpha, \beta$ and $s=\uparrow, \downarrow$ are layer, orbital and spin degrees of freedom, respectively, $\Psi_{E_{i},l\sigma s \textbf{x}}$  are the corresponding components of the eigenstate of the system. For comparison, we calculated the LDOS at the corner ($x\!=\!1, y\!=\!1$) and the edge ($x\!=\!1, y\!=\!L_{y}/2$) of a square sample [the same as the sample in Fig. 2(f) in the main text].

 Figures \ref{figS1}(a) and \ref{figS1}(b) illustrate the LDOS at the corner \textbf{x}$(x\!=\!1,y\!=\!1)$ and the edge \textbf{x}$(x\!=\!1,y\!=\!L_{y}/2)$ with respect to energy $E$ for different in-plane Zeeman field strength $h_{x}$. Obviously, when $h_{x}\!>\!h_{x}^{c}\approx 2.06$, we see that zero energy peaks develop in the LDOS at the sample corner. In contrast, the U-shape LDOS of the edge retain even for $h_x>h_x^c$. Therefore, the corner states in this system can be determined by the Zeeman field dependent zero bias peaks of different conductance ($dI/dV$) in STM measurements. Note that, the peak splitting for $h_x=2.1$ is because of the size effect that causes the overlap of wavefunctions of corner states. The corner states become more localized and the splitting vanishes as $h_x$ increases.

\end{document}